\documentclass[runningheads]{llncs}

\usepackage{amsmath,amssymb,amsfonts}
\usepackage{algorithmic}
\usepackage{graphicx}
\usepackage{textcomp}
\usepackage[T1]{fontenc}
\usepackage{xcolor}
\usepackage{hyperref} 
\def\BibTeX{{\rm B\kern-.05em{\sc i\kern-.025em b}\kern-.08em
    T\kern-.1667em\lower.7ex\hbox{E}\kern-.125emX}}

\begin{document}

\title{TracE2E: Easily Deployable Middleware for Decentralized Data Traceability}

 \author{Daniel Pressensé\and
 Elisavet Kozyri}
 \authorrunning{D. Pressensé et al.}

 \institute{UiT The Arctic University of Norway\\
 \email{\{daniel.pressense,elisavet.kozyri\}@uit.no}}

\maketitle

\begin{abstract}
This paper presents TracE2E, a middleware written in Rust,
that can provide both data explainability and compliance across multiple nodes. 
By mediating inputs and outputs of processes, TracE2E 
records provenance information 
and enforces data-protection policies (e.g., confidentiality, integrity) that depend on the recorded provenance.
Unlike existing approaches that necessitate substantial application modifications, TracE2E is designed for easy integration into existing and future applications through a wrapper of the Rust standard library's I/O module. 
We describe how TracE2E consistently records provenance information across nodes, and we demonstrate how the compliance layer of TracE2E can accommodate the enforcement of multiple policies.

\end{abstract}

\section{Introduction}
In today's data-processing systems, there is an increasing need for explainability and compliance.
Data regulations demand practices that can enable users understand how their data
is processed; how the final digital products they consume are created, and what
the origins of these products are.
Data regulations also prescribe restrictions on how data is allowed to be used: who is allowed to read and manipulate information that is directly or indirectly derived
from data provided by users.

A common denominator for faithfully supporting explainability and compliance is the
ability to trace data throughout its lifetime: from the moment of capture, to the processing stages, to the moment of consumption. 
Data traceability is usually accomplished by associating data with metadata that captures information about how this data was computed.

Given that data-processing applications span multiple nodes, data traceability should be supported across all nodes as well.
There are approaches that support distributed traceability~\cite{hutchison_safeweb_2011,lomotey_wearable_2017,mont_towards_2003,rongna_provenancebased_2021}.
However, these approaches require extensive modifications to existing applications in order to be deployed.

For a traceability infrastructure to be compelling, it needs to be easily integrated into
existing and future applications.
In particular, the logic of the application should be oblivious to the existence of a traceability infrastructure.
Consequently, a considerable number of proposed traceability systems have been developed at the operating system level~\cite{vandebogart_labels_2007,muniswamy-reddy_provenance-aware_2006,muniswamy-reddy_layering_2009,zeldovich_making_2011,pasquier_practical_2017,pasquier_runtime_2018}.
However, these systems focus on a local host approach, and thus, they support one node.
Operating system approaches that can support distributed applications \cite{zeldovich_securing_nodate,cheng_abstractions_2012} have had a compliance focus, and thus, they cannot be directly used to serve more general traceability purposes (e.g., explainability).

We aim to enforce complex reactive compliance policies based on provenance in distributed contexts, where multiple entities (middlewares) may decide whether to allow or deny data flows. More lightweight kernel-level solutions like eBPF are not suitable for this purpose. While eBPF can observe specific events in kernel space, it cannot report them to user-space for decision-making involving remote entities, as the kernel cannot be made to wait for feedback from user-space.

A traceability infrastructure is also expected to offer consistency between what is really happening in the system
and what is recorded to the associated metadata.
For traceability systems that aim to support multiple nodes, consistency becomes even more challenging to address.
Although all traceability tools are expected to provide a consistency level, the accompanied papers have not explicitly surfaced the involved intricacies and the adopted solutions.
We think surfacing the consistency problem in the context of data traceability, together with technical details 
about how this problem could be addressed, is important for future developers both in academia and in industry.

This paper presents a data-traceability tool, \emph{TracE2E}, that is
designed and implemented as a middleware and library in Rust.
It traces data as it flows between files and processes, by intercepting I/O actions that involve these entities.
TracE2E can support explainability, because its traceability layer records the provenance of each entity.
Our tool supports data traceability across multiple nodes.
We give a detailed description of how TracE2E consistently records traceability information across these nodes.
The front-end of TracE2E is a wrapper of the Rust standard library (I/O module),
and thus, it can be easily deployed by existing and future applications
 of the system entities.
TraceE2E can also support compliance as a layer built on top of traceability. 
According to our proposed design, one common traceability layer
can be used to potentially support different compliance layers built on top.
We are not aware of prior work that proposed an easily deployable and decentralized traceability system that can support both explainability and compliance with respect to different policies.

TracE2E has been implemented\footnote{TracE2E source code: \href{https://anonymous.4open.science/r/trace2e-2256}{https://anonymous.4open.science/r/trace2e-2256}.} and it has been evaluated qualitatively and quantitatively.
We show the straightforward workflow and the minimal code modifications
needed for deploying TracE2E to common existing applications.
And we present performance measurements that can be used to understand
the overhead added by TracE2E.
The evaluation opens up promising directions for future research.

\section{Design}
    The primary goal of TracE2E is to support both explainability and compliance services across multiple nodes, allowing to enforce policies that take provenance information into consideration.
    To accomplish this, TracE2E provides decentralized traceability through a middleware-based solution. The inputs and outputs (I/O) of processes across multiple nodes are qualified as data flows, which are systematically mediated to ensure traceability. 

\begin{figure}[ht!]
        \centering
        \includegraphics[width=0.5\linewidth]{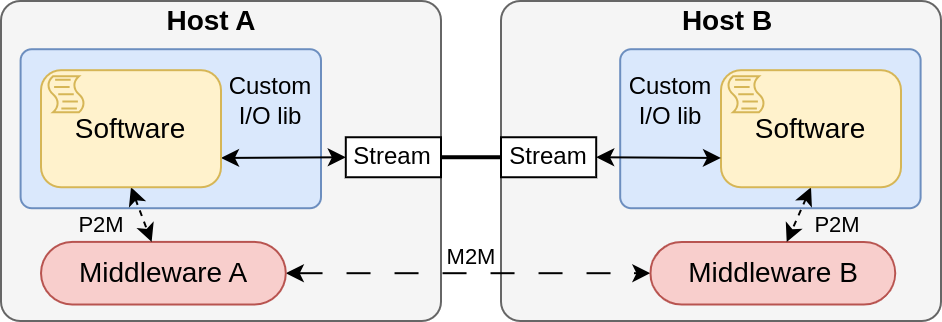}
        \caption{Overview of the TracE2E design}
        \label{fig:TracE2E_global_view}
    \end{figure}

    As represented in Figure \ref{fig:TracE2E_global_view}, the TracE2E framework for each host consists of a custom I/O library (in blue), wrapping the standard one, and a middleware (in red). The middleware components of each host interact with each other to provide end-to-end decentralized traceability. This approach enables decentralization and lightweight integration within existing systems.

    \subsection{Threat Model}
    For this work, an attacker is an honest but error-prone developer; 
    a developer that might inadvertently violate a data-protection policy or cannot fulfill an explainability request. So, TracE2E is deployed to prevent policy violations and to support explainability.
    It is assumed that all participating nodes running the TracE2E middleware behave honestly and strictly adhere to the proposed protocol.
    The problem of end-to-end traceability is challenging enough even under this threat model, which can explain the very limited
    research work on this problem.
    And there are practical scenarios where such a threat model 
    is relevant. 
    Consider for example, a company that owns a multi-node data-processing pipeline.
    Here, all nodes are trusted by the company to behave as expected. But the programmers of that company still benefit from the deployment of TracE2E to accomplish system-wide data traceability, instead of using ad-hoc approaches for explainability and compliance. 
    
    Our current solution is non-persistent, as traceability information is stored only in each local instance of the middleware. This means that if a middleware instance is restarted or fails, the traceability information it maintained is lost, and there is no persistent storage or recovery mechanism designed for the moment.  
    We see persistency as an orthogonal problem to the topic that this paper focuses on.

    \subsection{Provenance Layer}
        The provenance layer is responsible for tracking data flows between resources throughout the system. In our model, resources can be processes, files, or TCP streams, each uniquely identified within the system. A flow occurs when data is transferred between resources through read, write, or network operations, allowing the tracing of data movement across the system.
        
        For each resource, we maintain provenance information consisting of a list of resource identifiers. 
        When a flow between two resources occurs, the provenance of the destination resource is updated with the provenance of the source resource and the identifier of that source. For example, when a process $P1$ reads from a file $A$ and then writes to another file $B$, the provenance information of the file $B$ would include both the process $P1$ and the file $A$.

        The middleware maintains a mapping between resource identifiers and their associated provenance information. This mapping allows for the retrieval and update of provenance data as resources interact within the system. The provenance model provides two primary operations: (1) retrieving the current provenance information for a resource and (2) updating the provenance information when a flow occurs between resources.

        While more granular solutions exist for tracking data flows at finer levels (e.g., variable-level or instruction-level tracking), they do not serve our purposes and represent an orthogonal problem that can be incorporated in future work. Such fine-grained approaches require deeper program modifications beyond I/O library replacement. Given that the problem is already challenging enough, our contribution focuses on a generic provenance-based flow control model that achieves system-wide traceability through minimal infrastructure changes.

    \subsection{Provenance Recording Consistency}
        When a flow occurs from a source to a destination, there are two options to record provenance: synchronous and asynchronous. Under synchronous recording, the provenance of the destination is updated when the actual flow occurs. Under asynchronous recording, the provenance of the destination might be updated later than the time that the flow occurs. TracE2E employs synchronous recording of provenance, to ensure that the provenance information accurately represents the current state of the system at any moment.

\begin{figure}[ht!]
            \centering
            \includegraphics[width=0.45\linewidth]{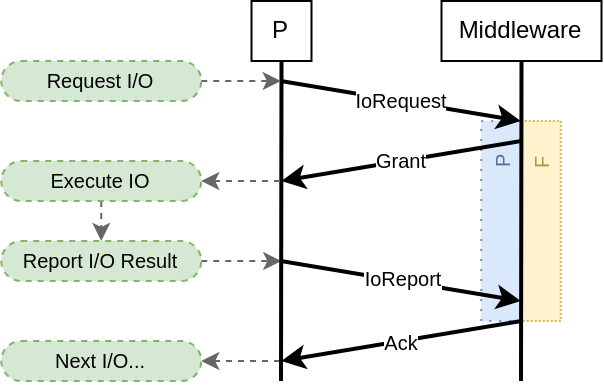}
            \caption{Processes to Middleware communication protocol (P2M), blue box represents the read reservation on source (Process P) and yellow one the write reservation on destination (File F) of the data flow}
            \label{fig:P2M-protocol}
        \end{figure}

        To accomplish synchronous recording of provenance, TracE2E implements a communication protocol between processes and the local middleware (P2M). This protocol treats each I/O operation as an atomic sequence of three steps: 1. Authorization; 2. Execution; 3. Reporting. As illustrated in Figure \ref{fig:P2M-protocol}, the protocol follows a specific message exchange pattern: First, the process sends an "IoRequest" message to the middleware and waits for a "Grant" response for Authorization; then the process performs the actual Execution of the I/O operation; finally, the process sends an "IoReport" message with the execution result to the middleware and receives an "Ack" to complete the Reporting phase. This structured sequence ensures that provenance information is consistently recorded at each step of data flow operations.
        
        Multiple processes that run in parallel on one node might cause flows that involve the same resource, potentially leading to race conditions when accessing the provenance of this resource. For example, when process $P$ reads from file $F$ and process $P'$ writes to $F$, the first flow involves the retrieval of $F$'s provenance and the second flow involves the update of $F$'s provenance. The potential concurrent access of $F$'s provenance might cause a race condition where the provenance information becomes inconsistent or incomplete.

        To prevent such race conditions, the naive solution would be to handle these flows and the respective provenance updates one by one using a global lock mechanism on the side of the middleware. However, this would create a dramatic bottleneck by serializing all I/O operations across the entire system, even when they do not involve correlated resources. Instead, TracE2E employs a more fine-grained locking mechanism based on per-resource reservation. In particular, TracE2E ensures that provenance accesses are performed atomically by implementing a Readers-Writers locking pattern, where multiple processes may read the provenance of a resource concurrently, but only one process at a time may update the provenance information. This approach allows concurrent operations on different resources while maintaining consistency for operations that affect the same resource.

        Before allowing a flow from a source to a destination, the provenance of the source resource is reserved with a read lock, the provenance of the destination resource is reserved with a write lock. After the flow (i.e., corresponding I/O operation) is performed and the provenance of the destination is updated with the provenance of the source, the reservation of both resources is released. The blue and yellow areas in Figure \ref{fig:P2M-protocol} represent respectively the read and write reservations of process and file resources, ensuring atomic execution of the entire I/O sequence. This combination of atomic execution and synchronous recording guarantees that all processes observe the same sequence of events, thereby maintaining system-wide coherence in the traceability information.

        Because the locks of two resources need to be reserved every time a flow is performed, there might be a risk of deadlock. For instance, if the performance of one flow involves the reservation of resource A and then resource B, while another concurrent flow involves the reservation of resource B and then resource A, a circular wait condition might occur. TracE2E prevents this deadlock by imposing a fixed reservation order based on resource type.

\begin{figure}[ht!]
            \centering
            \includegraphics[width=0.8\linewidth]{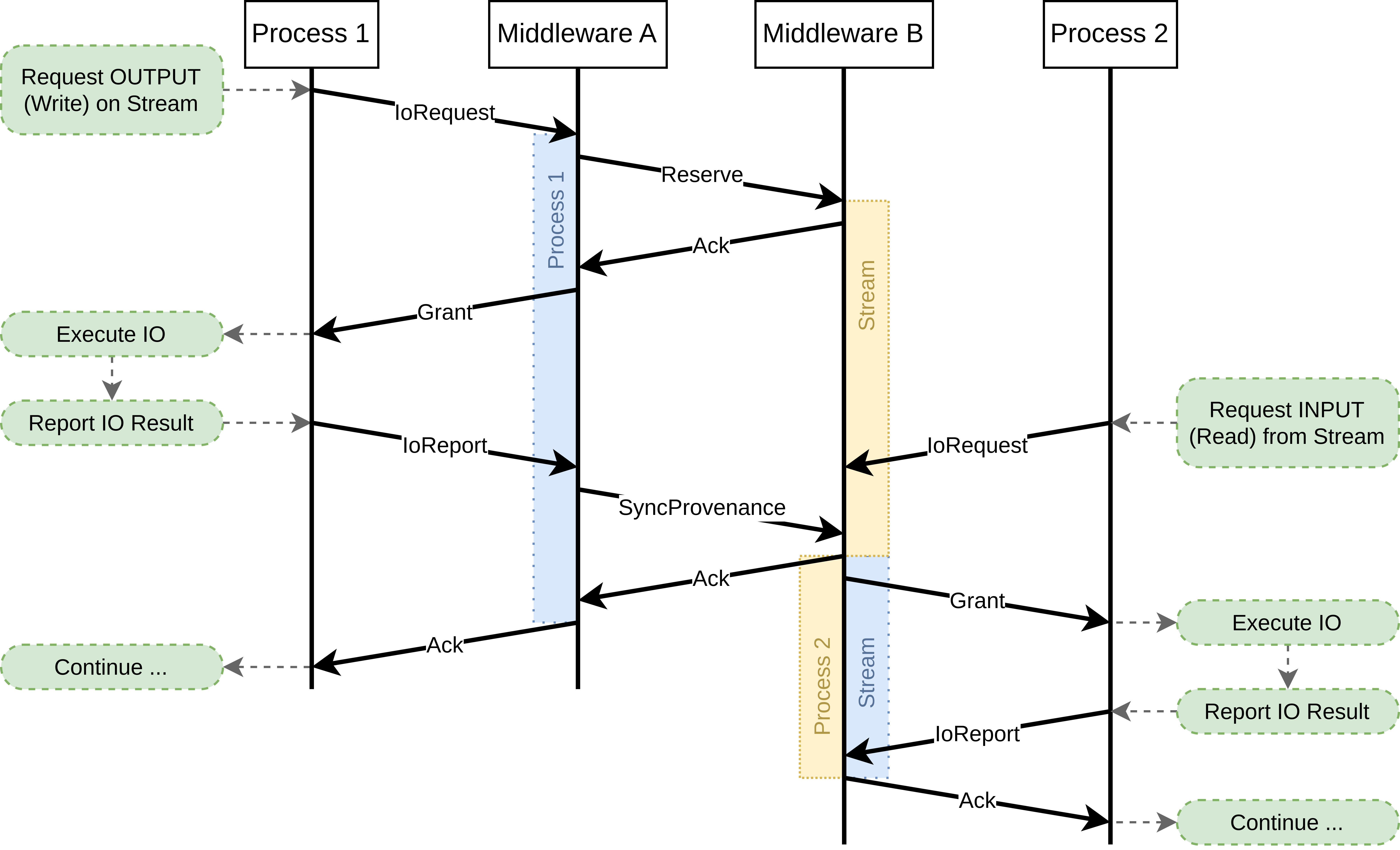}
            \caption{Middleware to Middleware (M2M) communication protocol enabling provenance recording in decentralized context. Blue and yellow boxes represent respectively the read and the write reservation of the resources involved in the data flows}
            \label{fig:M2M-protocol}
        \end{figure}

        TracE2E also supports synchronous and atomic recording of provenance when data flows across nodes. Consider a process $P1$ that requests to send data to a remote process $P2$ through a stream $S$. As illustrated in Figure \ref{fig:M2M-protocol}, the middlewares on both nodes coordinate to ensure consistent provenance recording using the Middleware-to-Middleware protocol (M2M). When $P1$ sends an "IoRequest" message to middleware A, the middleware first reserves the provenance of resource $P1$ locally, then sends a "Reserve" message to middleware B to reserve the remote side of stream $S$. Once middleware B acknowledges this reservation with an "Ack" message, middleware A sends a "Grant" message back to $P1$, completing the authorization phase. After $P1$ executes the write operation, it sends an "IoReport" message to middleware A, which then sends a "syncProvenance" message to middleware B to propagate the provenance update resulting from the data flow. The provenance of $S$ is updated with the provenance of $P1$, and an acknowledgment is sent back to $P1$, causing the release of $P1$ and $S$. Next, the provenance of resources $P2$ and $S$ is reserved, data is transferred from $S$ to $P2$, the provenance of $P2$ is updated with the provenance of $S$, and finally both resources are released.

        Consequently, for cross-node data flows, TracE2E treats TCP Stream resources similarly to files on each side. The M2M protocol extends the local P2M protocol by enabling coordination between middlewares. This communication protocol remains transparent to the processes, which continue to interact with their local middleware through the P2M protocol.\footnote{Even if TCP streams behave differently from files after a read operation: once read, it no longer holds the data, which would normally require removing the associated provenance information. However, since only one process holds the socket on each endpoint, residual out-of-date provenance from the stream will have no impact on the overall provenance state.}

    \subsection{Compliance Layer}
        Considerable amount of research has been focused on provenance-based approaches for policy enforcement. For instance, information flow control systems like Asbestos~\cite{vandebogart_labels_2007} and HiStar~\cite{zeldovich_making_2011} utilize process labels to track data contamination, while systems like SafeWeb~\cite{hutchison_safeweb_2011} leverage provenance information to enforce data flow constraints in web applications. These examples demonstrate how provenance information can serve as a foundation for implementing security policies.

        Our middleware framework extends this approach by providing a flexible infrastructure for policy enforcement for distributed systems. The compliance layer operates as a distinct component that leverages the underlying provenance information maintained by the provenance layer. This separation of concerns allows for the implementation of various security policies without modifying the core provenance tracking mechanism.

        To demonstrate these concepts, we use two simple provenance-based policies. The \emph{local confidentiality} policy ensures that sensitive data remains within the local host by examining the provenance of any data attempting to flow to external resources. Similarly, the \emph{local integrity} policy prevents contamination of protected resources by external data by analyzing the provenance of incoming data flows. 
        We show later that the traceability layer of TracE2E has been used to enforce both these policies.

        A critical aspect of our design is the timing of compliance checks. After resources are reserved but before the actual data flow occurs, the compliance layer evaluates whether the proposed operation adheres to all applicable policies. This checkpoint ensures that policy violations are prevented rather than merely detected. For example, before a process can write data derived from a local file to a network stream, the compliance layer verifies that such an operation is permitted by the desirable policies.

        The effectiveness of compliance enforcement fundamentally depends on the consistency of the underlying provenance information. Without consistency in provenance recording, policy decisions might be made based on incomplete or inconsistent provenance information, potentially leading to security violations. Consider a scenario where two processes simultaneously attempt to access a sensitive resource: if provenance updates are not properly synchronized, one process might gain unauthorized access before the system records the provenance information that would have prevented it. This illustrates why our middleware's provenance consistency, as described in the previous section, is essential for reliable policy enforcement.

\section{Implementation}

    TracE2E is built entirely using Rust, which was chosen for several key advantages. Rust provides memory safety guarantees without garbage collection and its ownership model prevents memory-related bugs at compile time. Rust's zero-cost abstractions enable high-level interfaces while maintaining performance comparable to lower-level languages, minimizing overhead in our traceability solution. Additionally, Rust's strong type system and concurrency model are well-suited for implementing the P2M and M2M communication protocols and resource reservation mechanisms required for consistency across distributed nodes. The language's growing ecosystem of libraries for systems programming and asynchronous operations (e.g., Tokio) provides the necessary building blocks for our traceability framework to integrate seamlessly with existing applications.

    Our TracE2E solution consists of two main components, the middleware with an instance running on each node (red boxes on the Figure \ref{fig:TracE2E_global_view}), and the custom I/O library, which all the programs to be traced are compiled with (blue boxes on the fig. \ref{fig:TracE2E_global_view}).

\begin{figure}[htbp]
      \centering
      \begin{minipage}[b]{0.4\textwidth}
        \centering
        \includegraphics[width=\textwidth]{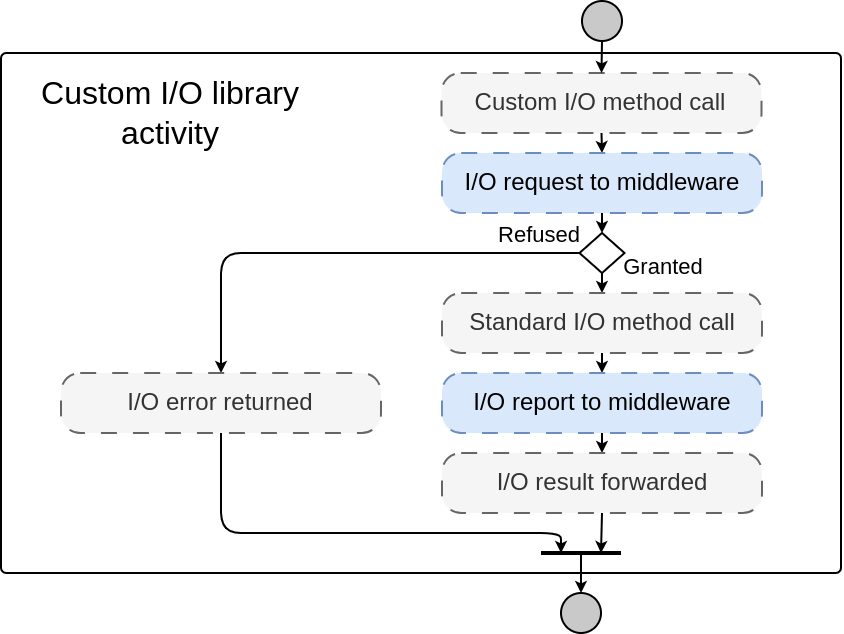}
        \caption{Custom I/O library activity. The grey boxes represent the execution of standard library components, and the blues ones the interaction checkpoints with the middleware.}
        \label{fig:custom_lib_activity}
      \end{minipage}
      \hfill
      \begin{minipage}[b]{0.55\textwidth}
        \centering
        \includegraphics[width=0.8\textwidth]{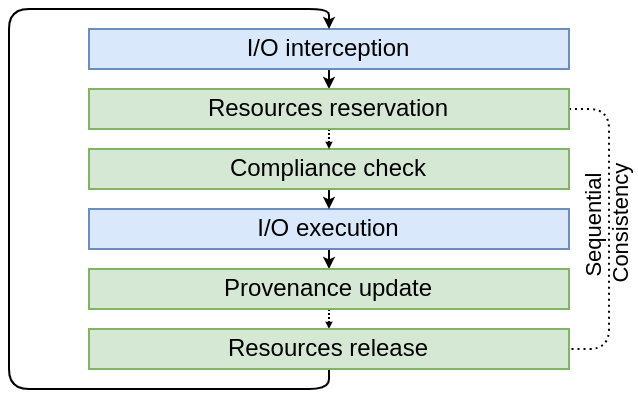}
        \caption{Traceability management stages. Custom I/O library handles blue boxes, while middleware manages green boxes. Solid arrows show P2M protocol communications, while dashed arrows represent internal middleware communications.}
        \label{fig:traceability_stages}
      \end{minipage}
    \end{figure}

    The custom I/O Library provides an interface that mirrors the types and method names of the standard Rust I/O library, thus requiring minimal modifications to existing applications. Additionally, it implements the P2M communication protocol. For each method that could initiate a data flow between two resources (i.e., Process and File or TCP Stream), the library employs an algorithm grounded in the design of the P2M protocol. This algorithm involves three main steps: first, a request is sent to the middleware to reserve the resources affected by the data flow and to ensure compliance for the flow; second, if the flow is granted, the standard method is executed; finally, the result of the standard method is reported back to the middleware (cf. Figure \ref{fig:custom_lib_activity}).

    Two variants of the custom I/O libraries are implemented: one for the Rust standard library and another for the Tokio library, which is the most popular asynchronous runtime for Rust. To achieve compatibility with any existing applications written in Rust within the context of our traceability framework, it is sufficient to replace the references to the existing I/O library with the customized version and then compile the application.

    The middleware is a foundational element of the TracE2E framework, operating as a persistent background service on each node to ensure system-wide traceability. Its key function is to mediate I/O operations, enforcing compliance policies while systematically recording provenance information. By integrating with the custom I/O library, the middleware guarantees that data flows between processes and resources are continuously mediated.

    The middleware employs a modular and loosely coupled design, promoting extensibility and maintainability. At its core, the traceability module orchestrates the resource reservation mechanism while interacting with two specialized layers: the provenance layer, by maintaining a reference system capturing resources interferences, and the compliance layer, which evaluates and enforces security policies taking into account information from provenance layer. Additionally, the middleware incorporates auxiliary modules that handle external communications, implementing P2M and M2M protocols that help with the coordination of the traceability management stages (cf. Figure \ref{fig:traceability_stages}) between the middleware and the processes running with the custom I/O library.

    \subsection{Global Resource Identification System}
        In order to provide decentralized traceability, each real-world resource (e.g., files, processes, TCP streams) is mapped to an in-memory object in the middleware of each node. Unique global identifiers are assigned to each resource to maintain consistency and uniqueness across nodes.
        
        The identifiers are structured to include the middleware ID and resource-specific information. For processes, this includes the process ID (PID) and start time to maintain runtime uniqueness despite PID recycling. Files are identified by their absolute path, while TCP streams use both local and peer socket information.

        The pattern for representing these identifiers follows a URI-like format, ensuring that each resource has a globally unique identifier within the distributed system.

    \subsection{Resource Label Structure}
        The TracE2E framework’s crucial traceability component is its label structure. Labels associate resources with information about their provenance and compliance. Building on the global resource identification system, the middleware maintains an in-memory mapping where each resource identifier maps to its corresponding label object.
            
        The labels structure is implemented as a Rust struct that encapsulates two main fields for provenance and compliance. The labels structure methods are grouped into two purpose-based interfaces for provenance and compliance and can operate on both fields of the labels highlighting the connection between our traceability layers.
        
        The provenance field is implemented as a list of labels objects, representing the complete provenance of resources that have influenced the data. This list grows dynamically as data flows through the system, with each flow operation appending new provenance information. The implementation follows a value-based propagation approach rather than reference tracking, meaning that during a data flow, the complete provenance collection from the source is copied and merged into the destination's provenance. This approach ensures that each resource maintains a self-contained and comprehensive record of its entire provenance, facilitating decentralized traceability without requiring centralized coordination.
        
        The compliance field is implemented as a struct containing boolean fields (flags), one for each supported policy (e.g., local integrity, local confidentiality). When a flag is set to true, the corresponding policy is enforced for that resource.
        
        By integrating both provenance and compliance components within a single label structure, TracE2E ensures that compliance policies and data provenance of each resource remain bonded throughout the system. This is crucial for maintaining cross-boundary consistency in distributed environments, where data flows may span multiple nodes.
  
    \subsection{External Communication Interfaces}
        The middleware provides gRPC-based communication interfaces for the simple and straightforward implementation of  P2M and M2M protocols through the Tonic Rust library. The implemented protocol structures refer to the resources involved in data flows using the global identification system previously described. The use of this technology is primarily motivated by research purposes, as it offers simple serialization features through Protobuf and high abstraction level for the data structures used in the TracE2E framework communications.

    \subsection{Resources Reservation}
        At the heart of the middleware lies the core traceability module, which is responsible for the consistency of resource representation while managing the provenance and compliance layers.
        
        This module employs message-passing to provide a flexible internal asynchronous API. A single thread is dedicated to handle this API and prevent race conditions. In fact this thread leverages the Rust's strict memory ownership model, by owning the hash map linking identifiers and the corresponding label objects. Each label object is guarded behind a RwLock structure that provides write and read lock features described in the design section.
        
        Upon each I/O request received by the gRPC interface, the core traceability module is called through message passing. This causes a dedicated thread to be spawned. This new thread is responsible for reserving the labels objects corresponding to the source and destination resource identifiers. To achieve the reservations, the thread is provided with a cloned reference to a RwLock objects safe guarding the label objects. The thread acquires a read lock for the source and a write lock for the destination. Once both locks are acquired, the traceability enforcement (see \ref{sec:traceability}) is performed with the reserved labels objects. If the I/O operation is allowed, the thread returns a grant id to the gRPC interface and creates a callback channel to yield the control back to the main thread. Later when the gRPC interface receives the report of the I/O operation, it triggers the traceability update (see \ref{sec:traceability}) and then releases the locks through the Rust scoped memory ownership system.
        
        To avoid the possible deadlocks described in the design section the label object associated to process identifier is always reserved first. In addition, a timeout is configured on each reservation thread. This guarantees that the reservation will not persist indefinitely in case of a failure in one of the traceability management stages. Such a failure would result in the definitive lock of the concerned resources.

    \subsection{Traceability}
    \label{sec:traceability}
        Following successful resource reservation, traceability enforcement operates as a two-phase atomic process within the same thread that manages resource locks, preventing race conditions and ensuring consistency.

        The first phase performs compliance checking through a flexible architecture centered around a main method applied to the destination resource's label object, using the source resource's label as an argument. This method evaluates comprehensive metadata including resource type, provenance, and location to return a boolean decision. The implementation supports both internal and external policy enforcers, employing lazy evaluation that terminates immediately upon detecting violations for enhanced efficiency and security. Internal enforcers demonstrate resource-type and provenance-based policies for local integrity and confidentiality enforcement.

        Upon successful compliance validation and I/O operation completion, the second phase executes provenance update. Each resource label structure maintains dedicated provenance fields recording resource interference. Source provenance information propagates to the destination resource. This two-phase approach maintains sequential consistency for the record of data flows, conveying both resource identifiers and traceability information to ensure complete explainability and compliance of subsequent operations.

        Figure \ref{fig:provenance_management_activity} in Appendix shows the detailed provenance management activity.

\section{Evaluation}
    The evaluation of TracE2E is conducted to demonstrate its effectiveness in enforcing compliance policies within a decentralized infrastructure while ensuring usability for the user. A proof of concept is developed based on a plausible real-world scenario involving data flows between multiple files and processes over multiple nodes. The primary objectives of this section are to demonstrate the enforcement of provenance-based compliance policies, show ease of integration and usage of the traceability framework, and evaluate overall system performance.

    \subsection{Experimental Setup}
        Our evaluation utilizes example programs provided by Hyper.rs, a popular Rust HTTP library built on the Tokio asynchronous runtime. We leave these examples largely unmodified to demonstrate how little work is required to make existing Rust applications support the TracE2E framework. This approach showcases TracE2E's seamless integration with our patched Tokio library in real-world applications.
        
        For performance measurements, we developed simple benchmark programs that perform sequences of read and write operations on Files and TCPStreams using both the standard library and our modified I/O library. Each benchmark was executed 1000 times to ensure statistical significance. The overhead for each component of the framework was determined using the Rust tracing library, which allowed us to checkpoint the beginning and end timestamps of each component's execution, providing a detailed breakdown of performance costs.

        The evaluation was conducted on a laptop with an Intel Core i7-1260P processor and 32GB of RAM. For multi-node scenarios, we employed a Docker Compose infrastructure to simulate distributed environments and ensure reproducibility\footnote{All presented experiments are reproducible with the resources available in the demo directory of TracE2E git repository.}.

    \subsection{Deployment Evaluation}
\begin{figure}[ht!]
            \centering
            \includegraphics[width=0.4\linewidth]{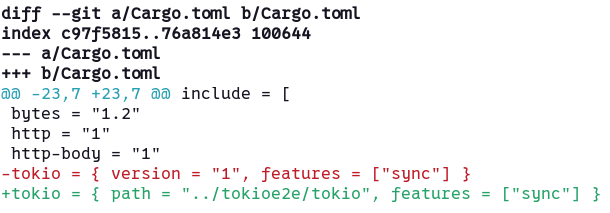}
            \caption{Example TracE2E patch applied to the Cargo file of an existing Rust project}
            \label{fig:cargo_diff}
        \end{figure}

        TracE2E provides customized I/O libraries: a standard library wrapper (stde2e) and patches for the Tokio asynchronous runtime. Integration with existing applications requires minimal effort—in the case of Hyper.rs applications, we simply modified the Cargo.toml file to replace the original Tokio dependency with our modified version, as shown in Figure \ref{fig:cargo_diff}. Notably, no changes to the application source code were required. This approach enables seamless integration with any Rust application, significantly reducing adoption barriers.

    \subsection{Policy Enforcement Evaluation}
        \subsubsection{Confidentiality Policy}
\begin{figure}[ht!]
                \centering
                \includegraphics[width=0.7\linewidth]{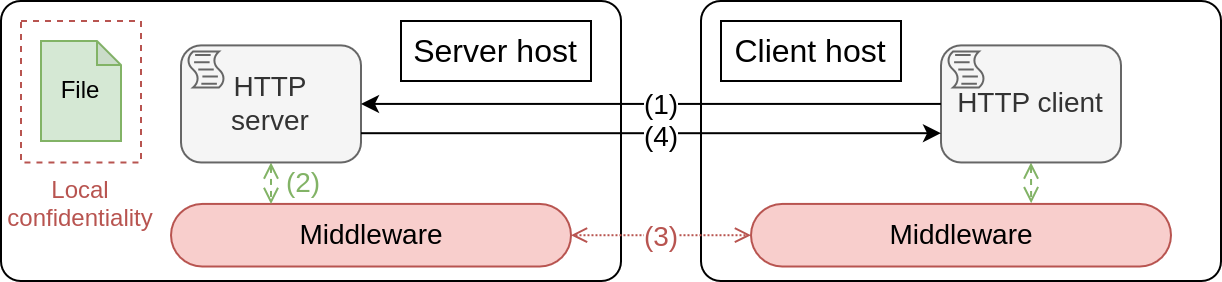}
                \caption{Local confidentiality policy enforcement workflow: (1) HTTP request from remote client, (2) P2M protocol consultation, (3) M2M coordination between middlewares, (4) policy decision detection for confidentiality violation, (5) empty response.}
                \label{fig:example_hyper_local_confidentiality}
            \end{figure}

            In our evaluation setup, shown in Figure \ref{fig:example_hyper_local_confidentiality}, we demonstrate how TracE2E enforces confidentiality policies across distributed nodes. The scenario involves a web server process hosting sensitive HTML files, a remote client process attempting to access these files, and TracE2E middleware on both nodes coordinating policy enforcement through provenance tracking. We use Docker containers to simulate this distributed system, with separate containers representing the client machine and server machine, each running its own TracE2E middleware instance.

            The local confidentiality policy prevents sensitive data from flowing to external resources. Without TracE2E, a misconfigured web server would freely serve protected HTML files to external clients. With TracE2E enabled, the enforcement workflow proceeds as follows: (1) the remote client sends an HTTP request for the sensitive file, (2) the web server consults its local middleware via the P2M protocol, (3) the middlewares coordinate via the M2M protocol to verify the cross-node data flow, (4) the middleware identifies that serving the confidentiality-protected file to an external client violates the policy, and (5) the server returns an empty response instead of the sensitive content. This enforcement occurs transparently to the application, which continues to use standard I/O interfaces while the middleware mediates access according to established policies.

            The policy enforcement is provenance-based because it tracks the complete history of how the sensitive data was created and modified, and it prevents data leakage even if the data has been transformed or combined with other data.

        \subsubsection{Integrity Policy}

\begin{figure}[ht!]
                \centering
                \includegraphics[width=0.7\linewidth]{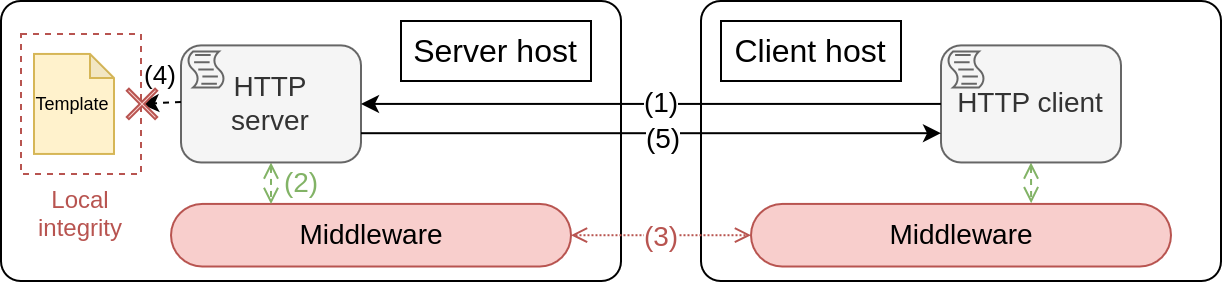}
                \caption{Local integrity policy enforcement workflow: (1) HTTP request from remote client, (2) P2M protocol consultation, (3) M2M coordination between middlewares, (4) provenance-based policy decision detecting integrity violation, (5) error response.}
                \label{fig:example_hyper_local_integrity}
            \end{figure}

            Figure \ref{fig:example_hyper_local_integrity} demonstrates TracE2E's enforcement of integrity policies in a distributed setting. The evaluation setup features a template HTML file restricted to local modifications, a remote client process attempting to modify it, and TracE2E middleware on both nodes coordinating to prevent unauthorized external modifications through provenance analysis.

            The local integrity policy blocks external processes from modifying protected data. Without this policy, remote modification attempts succeed normally, potentially compromising data integrity. With TracE2E enabled, the enforcement workflow operates as follows: (1) the remote client sends an HTTP PUT request with modification data, (2) the server process consults its local middleware via the P2M protocol, (3) the middlewares coordinate via the M2M protocol to analyze the modification source, (4) the middleware identifies that the modification originates from an external process, violating the integrity policy, and (5) the server returns a 403 Forbidden response, blocking the unauthorized modification.

            The policy enforcement is provenance-based because it tracks the chain of processes that have influenced the modification to prevent data modification by external processes. The middleware's decision to block the template file write relies on analyzing the provenance information of the HTTP server process. As it received a request from a remote, it will convey a reference to remote resource in its provenance information, altering the template file would result in integrity violation.

    \subsection{Performance Evaluation}
\begin{figure}[htbp]
          \centering
          \begin{minipage}[b]{0.49\textwidth}
            \centering
            \includegraphics[width=\textwidth]{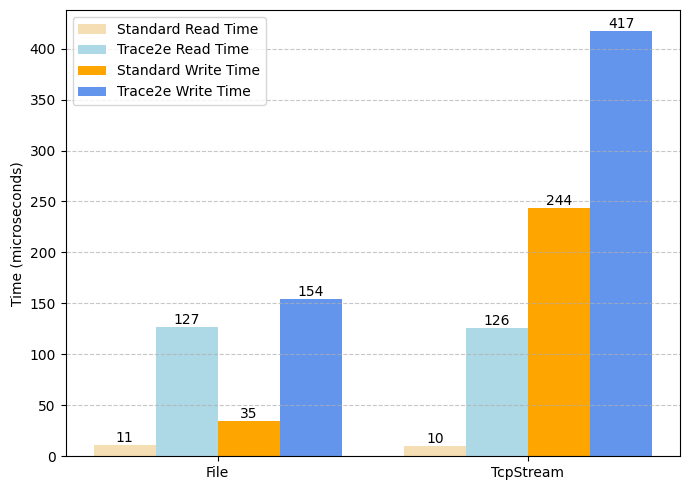}
            \caption{Comparison of I/O mean performance: Standard vs TracE2E}
            \label{fig:TracE2E_overhead}
          \end{minipage}
          \hfill
          \begin{minipage}[b]{0.49\textwidth}
            \centering
            \includegraphics[width=\textwidth]{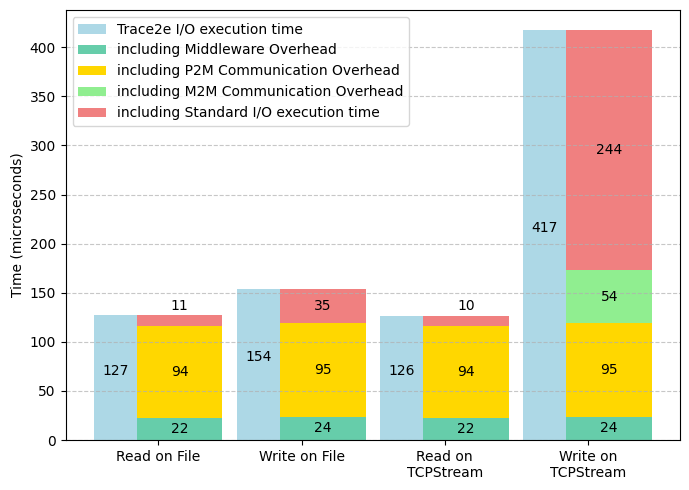}
            \caption{TracE2E I/O execution overhead breakdown}
            \label{fig:TracE2E_overhead_breakdown}
          \end{minipage}
        \end{figure}

        Our performance evaluation reveals that TracE2E introduces significant overhead to I/O operations. As shown in Figure \ref{fig:TracE2E_overhead}, standard-library read-file operations complete in around 11$\mu s$ and write-file operations in 35$\mu s$, while TracE2E-mediated read-file and write-file operations take considerably longer, respectively 127$\mu s$ and 154$\mu s$. Similar overhead magnitude emerge for TCPStream operations in our evaluation\footnote{These measurements only consider the overhead around the method call. When a write method returns, it does not mean the data is synced to disk. If the sync time was considered, the write overhead imposed by TracE2E would be negligible}.

        Figure \ref{fig:TracE2E_overhead_breakdown} breaks down the sources of this overhead. The major part comes from process-to-middleware (P2M) and middleware-to-middleware (M2M) communication protocols based on the gRPC framework, which ensure consistency in provenance recording and policy enforcement. This overhead is constant per I/O operation, meaning that the relative impact decreases as the standard I/O operation duration increases. Applications performing many small, frequent I/O operations experience greater relative performance impact than those with fewer, larger operations.

        Interestingly, even by taking into account M2M protocol overhead, the relative overhead for TCPStream operations in decentralized contexts is comparable to that for File operations, demonstrating TracE2E's viability for distributed systems. These performance measurements highlight the trade-off between comprehensive traceability guarantees and computational efficiency, a consideration that must be weighed based on the specific security and compliance requirements of the application.

\section{Related Work}
    Kernel-level solutions like CamFlow~\cite{pasquier_practical_2017}, CamQuery~\cite{pasquier_runtime_2018}, and Linux Provenance Modules~\cite{bates_trustworthy_2015} provide comprehensive provenance capture with minimal overhead but lack distributed capabilities and require invasive kernel modifications. Unlike these approaches, TracE2E operates as middleware without kernel dependencies, enabling lightweight deployment across diverse distributed systems.

    Operating system approaches including Asbestos~\cite{vandebogart_labels_2007}, HiStar~\cite{zeldovich_making_2011}, and Aeolus~\cite{cheng_abstractions_2012} use process labeling for data contamination tracking, while PASS~\cite{muniswamy-reddy_provenance-aware_2006,muniswamy-reddy_layering_2009} focuses on provenance-aware storage systems. These solutions require substantial kernel modifications. Decentralized IFC models~\cite{bacon_information_2014} and sticky policy frameworks~\cite{mont_towards_2003} address cloud security concerns but lack practical implementation for general-purpose traceability. Middleware solutions like SafeWeb~\cite{hutchison_safeweb_2011} and IoT health systems~\cite{lomotey_wearable_2017} provide distributed traceability but require extensive application re-engineering. TracE2E minimizes integration effort by requiring only recompilation with our custom I/O library.

    Policy enforcement frameworks using formal methods~\cite{basin_monitoring_2015,hublet_proactive_2024,hublet_user-controlled_2024}, decentralized policy propagation~\cite{zhao_automated_2021,zhao_draid_2021}, and distributed enforcement systems~\cite{elnikety_thoth_2016} demonstrate sophisticated policy specification capabilities but lack integrated provenance capture mechanisms. TracE2E provides both consistent provenance recording and policy enforcement in a unified framework.

    Blockchain approaches~\cite{dwivedi_blockchain-based_2021,datta_blockchain-based_2024,ali_secure_2018} for provenance provide integrity through smart contracts but target high-level data exchanges rather than system I/O operations, making them unsuitable for fine-grained process-level traceability that TracE2E addresses.

    Most existing solutions require substantial system modifications, making deployment challenging in established infrastructure. TracE2E addresses this gap by providing decentralized traceability with minimal integration requirements—middleware installation and application recompilation—while maintaining consistency guarantees across distributed nodes.

\section{Conclusion}
    This paper presented TracE2E, a middleware solution for decentralized data traceability that addresses the challenge of tracking data flows across multiple nodes while maintaining consistency and enabling policy enforcement. TracE2E's key contribution lies in its minimal integration requirements—applications need only recompile with our custom I/O library—making it easily deployable in existing distributed systems.

    Our evaluation demonstrated TracE2E's ability in enforcing provenance-based policies such as local confidentiality and integrity constraints. The synchronous recording mechanism ensures consistency across nodes through fine-grained resource locking, while the modular design supports extensible compliance layers built atop the traceability foundation.

    Future work includes exploring more scalable consistency models for larger distributed environments, implementing "by-reference" policy propagation models to enable dynamic policy modifications, and integrating with real-time compliance frameworks for richer policy enforcement capabilities. These enhancements will extend TracE2E's applicability to increasingly complex regulatory and scalability requirements.

\section*{Acknowlegements}
For style refinement and translation, we utilized DeepL, and ChatGPT (GPT-4o) with the prompt "You are a researcher. Improve the following sentence to adopt an academic style..."


\bibliographystyle{plain}
\bibliography{main}
\newpage
\section*{Appendix}

\begin{figure}[ht!]
    \centering
    \includegraphics[width=0.6\linewidth]{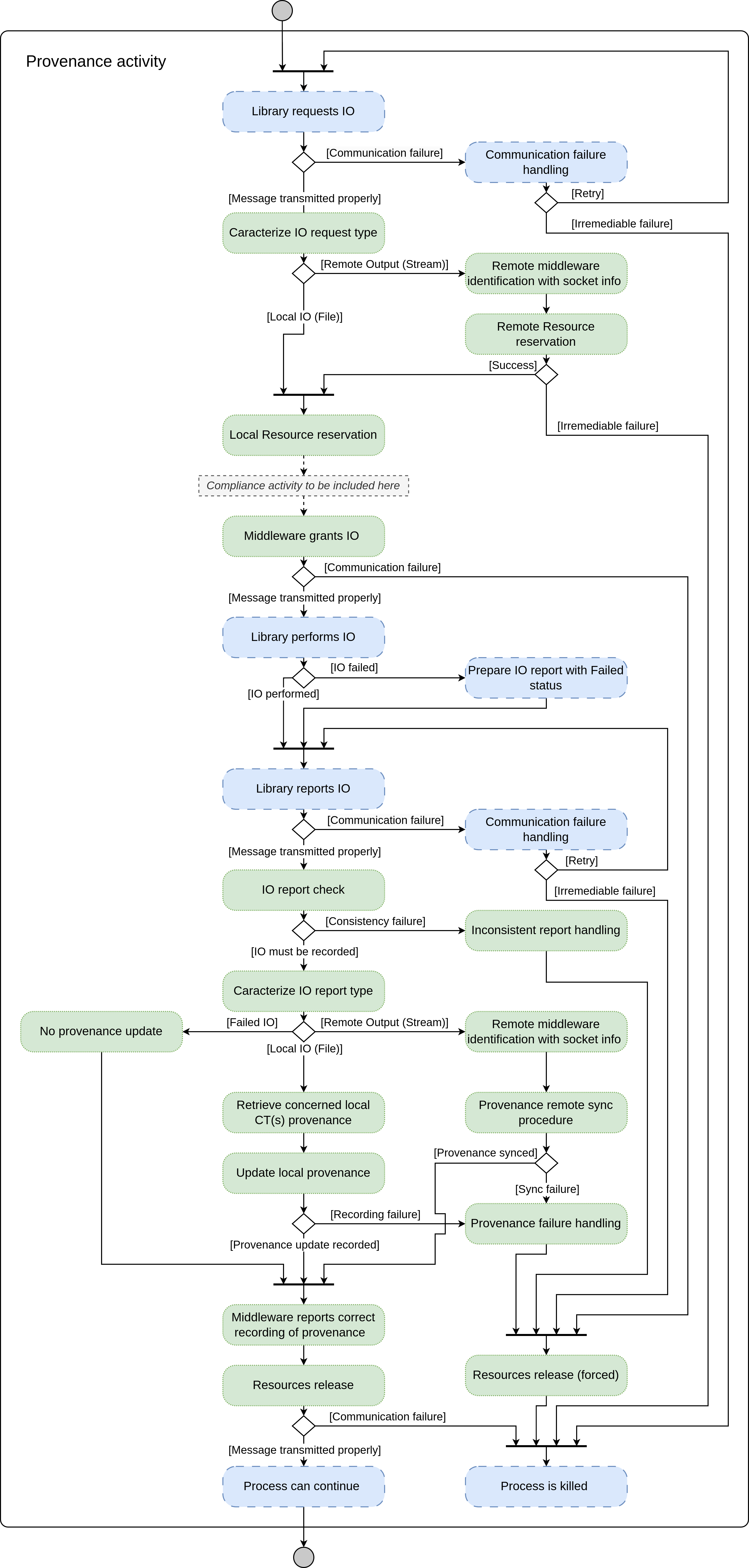}
    \caption{Detailed provenance management activity. The blue and green boxes, respectively, represent the actions performed by the custom I/O library and the middleware.}
    \label{fig:provenance_management_activity}
\end{figure}

\end{document}